\documentclass{tran-l}

 \theoremstyle{definition}
 
 \theoremstyle{remark}
 
 \newcommand{\bra}[1]{\left\langle#1\right\vert}
\newcommand{\ket}[1]{\left\vert#1\right\rangle}
\begin{document}

\title[Superluminality in Quantum Theory]
 {Superluminality in Quantum Theory}

\author{Giovanni Andrea Fantasia }

\address{Dipartimento di Informatica , Universita' di Milano}

\email{fantasia@dsi.unimi.it}

\thanks{I would like to thank E.Angeleri, G.Degli Antoni, F.Thaheld, S.Srikanth, E.Recami for useful discussions
         Thank also STMicroelectronics for financial support}


\date{}
\dedicatory{}

\commby{}

%%% ----------------------------------------------------------------------

\begin{abstract}
In the present work we investigate the possibility of superluminal
information transmission in quantum theory. We give  simple and
general arguments to prove  that the general structure (Hilbert's
space plus instantaneous state reduction) of the theory allows the
existence of superluminal communication. We discuss how this
relates with existing no-signalling theorems.
\end{abstract}

%%% ----------------------------------------------------------------------
\maketitle
%%% ----------------------------------------------------------------------

\section*{Introduction}
Since the EPR paradox appeared \cite{EPR35} the question of which
were the sense of non-locality in quantum theory was a very
fundamental question to address \cite{Red}. One main problem is to
know if quantum non-locality implies superluminal communication
between two separated parts A and B. Several arguments are given
to exclude this hypotheses \cite{Bohm} \cite{Ghi80} \cite{Shi}:
they form well-known no-signalling theorems. However these
theorems have all the common hypotheses that operators associated
with the two separated parts commute: [A,B]=0. We show possible
gedanken experiments which violate this hypotheses: in effect we
give two examples that prove that the general structure (Hilbert's
space plus instantaneous state reduction) of quantum theory allows
the existence of superluminal communication.

\section*{Superluminal communication}
In this section we propose two different protocols for
superluminal communication and we discuss how they circumvent
no-signalling theorems. Suppose that a quantum system, say a
particle, is in the state
\begin{equation}\label{particle}
\frac{1}{\sqrt{2}}(\ket{A}+\ket{B})
\end{equation}
where $\ket{A}$ and $\ket{B}$ represent two long distance
separated spatial localization of the particle. We assume that
trough all the duration of the protocol dispersion of the particle
will be not relevant.  So $\ket{A}$ represents the particle
localized in a finite volume in region A e so does $\ket{B}$. If
we make a measurement testing if particles is in the state
$\ket{A}$ this action is represented by the projector
$\ket{A}\bra{A}$ where $\bra{A}$ has the properties
\begin{equation}\label{proj}
\bra{A}A\rangle=1;\bra{A}B\rangle=0
\end{equation}
In this way observing the state (\ref{particle}) using
$\ket{A}\bra{A}$ will transform the state of the system in
$\ket{A}$ or in  with equal probability. Analogously we can define
a projector

\begin{equation}\label{projB}
\ket{B}\bra{B}
\end{equation} \\
Let construct a new projector
\begin{equation}\label{newproj}
\frac{1}{\sqrt{2}}\ket{A+B}\frac{1}{\sqrt{2}}\bra{A+B}
\end{equation}

defining
\begin{equation}\nonumber
\frac{1}{\sqrt{2}}\bra{A+B}=\frac{1}{\sqrt{2}}(\bra{A}+\bra{B})
\end{equation}
where $\frac{1}{\sqrt{2}}\bra{A+B}$  has the properties
\begin{align*}\frac{1}{\sqrt{2}}\bra{A+B}\frac{1}{\sqrt{2}}(A+B)\rangle=1&\\\frac{1}{\sqrt{2}}\bra{
A+B}\frac{1}{\sqrt{2}}(A-B)\rangle=0
\end{align*}
In this way observing the state $\ket{A}$ using projector
(\ref{newproj}) will transform the state of the system in
$\frac{1}{\sqrt{2}}(\ket{A}+\ket{B})$ or in
$\frac{1}{\sqrt{2}}(\ket{(A}-\ket{B})$ with equal probability. In
fact
$\ket{A}=\frac{1}{\sqrt{2}}(\frac{1}{\sqrt{2}}(\ket{A}+\ket{B})+\frac{1}{\sqrt{2}}(\ket{A}-\ket{B}))$
where $\frac{1}{\sqrt{2}}(\ket{A}+\ket{B})$ and
$\frac{1}{\sqrt{2}}(\ket{A}-\ket{B})$ are the eigen-vectors of
projector $\frac{1}{\sqrt{2}}\ket{A+B}\frac{1}{\sqrt{2}}\bra{A+B}$
. Making now a projection with $\ket{B}\bra{B}$ and whatever was
the state leaved by
$\frac{1}{\sqrt{2}}\ket{A+B}\frac{1}{\sqrt{2}}\bra{A+B}$ we will
find the particle in the state $\ket{B}$ and so with probability
$\frac{1}{2}$ we teleportate the particle from A to B. \\
Using a large collection of particle will permit us to raise the
probability that at least one particle reaches B close as we want
to $1$. \\
 Summarizing we have a particle in the localized state
$\ket{A}$, part A makes a measurement projecting onto
$\frac{1}{\sqrt{2}}(\ket{A}+\ket{B})$ then part B makes a second
measurement projecting onto $\ket{B}$ in a finite amount of time.
This protocol leads with probability $\frac{1}{2}$ to
teleportate particle from A to B. \\
The key feature of this protocol is the use of projector
(\ref{newproj}). We see that this projector doesn't commute with
the one associated with B (\ref{projB}). In fact
\begin{align*}
[\frac{1}{\sqrt{2}}\ket{A+B}\frac{1}{\sqrt{2}}\bra{A+B}\ket{B}\bra{B},
\ket{B}\bra{B}\frac{1}{\sqrt{2}}\ket{A+B}\frac{1}{\sqrt{2}}\bra{A+B}]=&\\
\frac{1}{2}\ket{A+B}\bra{B}-\frac{1}{2}\ket{B}\bra{A+B}=\frac{1}{2}(\ket{A}\bra{B}-\ket{B}\bra{A}\neq
0\end{align*} A protocol that uses only statistical correlations
between spin systems could be find in \cite{Ghi00}.\\
There is however  a physical objection that may be done to
measurements of the kind (\ref{newproj}). In effect to do this
measurement part A would physically operate on both region of
space A and B because this measurement test a non-local properties
of the system. We can test if the particle is in A or in B just
operating locally in A (or in B) but it is not clear if
measurement (\ref{newproj}) may be physically achieved by part A
operating locally in A (or in B) or even non-locally in A and in
B. Even if this operator is formally a well defined  self-adjoint
operator it is hard to imagine a physically (local or non-local)
realization of such a measurement. Anyway this problem doesn't
seems to affect the next example of superluminal communication.\\
Suppose now part A has a particle in the state $\ket{A}$. So the
particle is strictly localized   around A and probability that B
detect the particle is very near to  0. Part A performs (locally)
a precision measurement on the momentum of the particle. This
"collapse" the wave function from $\ket{A}$ to $\ket{A'}$. The new
state $\ket{A'}$, in momentum domain,  is strictly localized
around a random value $\lambda$ that is the result of precision
momentum measurement. Obviously, as indetermination relation
between momentum and position $2\Delta x\Delta p\geq\hbar$
requires, the new state $\ket{A'}$ is more spread in space than
old state $\ket{A}$ so that now probability that part B detects
the particle will be increased. So part A could send instantaneous
information to B. Again the protocol circumvent the no-signalling
theorems because measurement operators of part A and part B
doesn't commute. We do now some formal calculations to validate
this protocol.\\ Suppose part A  and B laying on a line. Part A is
located at the origin and B is located at a distance d from A.
Part B detects if the particle is located in the region [d-k,d+k].
So  the starting state $\ket{A}$ written in space domain is
\begin{equation}\label{stateA}
\ket{A(x)}=\frac{1}{\sqrt[4]{\pi\sigma^2}}\hspace{4pt}e^{-\frac{x^2}{2\sigma^2}}
\end{equation}
($\hbar=1$) with the variance small enough to let probability that
part B detect the particle near to 0
\begin{equation}\label{smallprob}
    \int^{d+k}_{d-k}\parallel\frac{1}{\sqrt[4]{\pi\sigma^2}}\hspace{4pt}e^{-\frac{x^2}{2\sigma^2}}\parallel^2 dx\approx
    0
\end{equation}
Fourier transform of state $\ket{A}$ will give us the wave
function written in momentum domain
\begin{equation}\label{momentum1}
F(\ket{A})=\ket{A(p)}=\sqrt[4]{\frac{\sigma^2}{\pi}}\hspace{4pt}e^{-\frac{\sigma^2p^2}{2}}
\end{equation}
After the precision momentum measurement, supposed we got the
random  value $p=\lambda$, the state $\ket{A'}$ written in
momentum domain will be
\begin{equation}\label{statexmom}
\ket{A'(p)}=\sqrt[4]{\frac{\overline{\sigma}^2}{\pi}}\hspace{4pt}e^{-\frac{\overline{\sigma}^2(p-\lambda)^2}{2}}
\end{equation} with $\overline{\sigma}\gg\sigma$.
In the space domain state $\ket{A'}$ will be
\begin{equation}\label{statexspc}
\ket{A'(x)}=F^{-1}(A'(p))=\frac{1}{\sqrt[4]{\pi\overline{\sigma}^2}}\hspace{4pt}e^{-\frac{x^2}{2\overline{\sigma}^2}-\frac{i\lambda
x}{2}}
\end{equation}
The probability that part B will detect the particle is now
\begin{equation}\label{prob2}
 \int^{d+k}_{d-k}\parallel \frac{1}{\sqrt[4]{\pi\overline{\sigma}^2}}\hspace{4pt}e^{-\frac{x^2}{2\overline{\sigma}^2}-\frac{i\lambda
x}{2}}\parallel^2 dx=\int^{d+k}_{d-k}
\frac{1}{\sqrt[2]{\pi\overline{\sigma}^2}}\hspace{4pt}e^{-\frac{x^2}{\overline{\sigma}^2}}dx>0
\end{equation} and we see that doesn't depend from the $\lambda$ measured
momentum.

\section*{Conclusion}
Even if superluminal velocity in unitary evolutions of quantum
system was already known \cite{Chi} \cite{Heg} we think that we
have here provides some clear examples of how hypotheses of
instantaneous wave collapse in quantum theory will permit
superluminal communication. This is not in conflict with existing
no-signalling theorems because we don't use commuting operators.
Our main conclusion is that quantum theory doesn't peacefully
coexist with special relativity and no-signalling requirements
will impose new postulates to quantum theory. In effect it turns
out that a sufficient condition for no-signalling \cite{Peres} is
$[A,B]=0$ where A are B are measurement operators associated with
part A and B but , as we have seen, we can imagine several
situations in which $[A,B]\neq 0$.

% ------------------------------------------------------------------------
%GATHER{Xbib.bib}   % For Gather Purpose Only
%GATHER{Paper.bbl}  % For Gather Purpose Only
%\bibliographystyle{amsplain}
%\bibliography{xbib}
\end{document}